\documentclass[%
 reprint,
 amsmath,amssymb,
 aps
]{revtex4}

\usepackage{graphicx}% Include figure files
\usepackage{dcolumn}% Align table columns on decimal point
\usepackage{bm}% bold math

\begin{document}
\preprint{APS/123-QED}
\title{
What connects ignition and deflagration?\\- On explosive transition of deflagration -
}% Force line breaks with \\
%\thanks{What connects ignition and deflagration?}%

\author{Youhi Morii}
 \email{morii@edyn.ifs.tohoku.ac.jp}
\author{Kaoru Maruta}%
 \email{maruta@ifs.tohoku.ac.jp}
\affiliation{%
Institute of Fluid Science, Tohoku University\\ 
2-1-1 Katahira, Aoba, Sendai, Miyagi, 980-8577, Japan\\
}%
\date{\today}% It is always \today, today,

\begin{abstract}
The relation between ignition and deflagration analysis theoretically shows that the temporal evolution of normalized fuel mass fraction and temperature in 0D homogeneous ignition are equivalent to the spatial evolution of those in a 1D laminar premixed flame at Lewis number of unity if a spatial-temporal transformation of the flame is applied.
  In addition, the degree of decrease in normalized fuel mass fraction in the preheat zone depends on the Lewis number, suggesting that ignition in the preheat zone near the explosive transition can only occur with Lewis numbers greater than unity.
\end{abstract}
\maketitle

%\tableofcontents

\section{Introduction}
Understanding of "extinctive transitions of deflagration" has been studied over the years.
Zeldovich \cite{Zeldovich1941} and Spalding \cite{Spalding1957} proposed a pioneering thermal theory of flame propagation limit in gas mixtures,
followed by Joulin and Clavin \cite{Joulin1976}.
%And Buckmaster applied asymptotic methods to 1D laminar premixed flame in the presence of heat loss to the tube wall and developed a thermal theory of the flame propagation limit　\cite{Buckmaster1976}.
And Buckmaster applied asymptotic methods to 1D laminar premixed flame in the presence of heat loss to the tube wall and developed a thermal theory of the flame propagation limit \cite{Buckmaster1976}.
Subsequently, numerous studies were continued on the "extinctive transitions of deflagration." Frankel and Sivashinsky \cite{Frankel1984} addressed the effects curvature on flames, and Ju et al. \cite{Ju1997} and Buckmaster \cite{Buckmaster1997} studied the combined effects of heat loss, Lewis number, and flame stretch on flames.
Series of studies on non-propagating spherical flame (e.g., Ronney et al. \cite{Ronney1994}), have indicated the existence of stable flame ball. Ju et al. summarized a global picture of extinctive transition of flames \cite{Ju2001}.
More recently, experimental and computational studies reported that the near extinction-limit flame interactions under low Lewis number conditions using a flame ball, a counterflow flame, and a planar flame \cite{Okuno2018, Tsunoda2022}.

On the other hand, "explosive transition of deflagration" has not been fully understood while it is important in safety engineering and practical combustor development.
Important and unresolved phenomena related to "explosive transition of deflagration" include detonation, which is important for safety engineering and necessary for understanding the mechanism of Type Ia supernova (SNIa), and knocking, which occurs in spark-ignition engines and is an obstacle to improving thermal efficiency.
For detonation phenomena, the structure of the detonation is described by the well-known ZND (Zeldovich-Neumann-Döring) model, but it is treated as an ignition problem behind the shock wave and does not consider the effect of flame propagation.
Therefore, in detonation problems, deflagration to detonation transition (DDT) is considered to be the appropriate target problem involving "explosive transition of deflagration."
Various studies on DDT, from ordinary flammable premixed gases to the SNIa were summarized by Oran \cite{Oran2015}.
While the causes of DDT were not fully understood, main mechanisms that have been obtained are thought to be interference among flame front, shock wave and local explosions where the "explosive transition of deflagration" play important role.

Another explosive event, knocking in SI engines is one of the typical phenomena related to "explosive transition of deflagration" which controls the improvement of thermal efficiency.
The Livengood-Wu integral based on ignition delay time \cite{Livengood1955} is a well-known method for predicting knocking.
The pioneering studies on knock onset mechanism by Zeldovich \cite{Zeldovich1980a} followed by Gu et al. \cite{Gu2003b} showed that the existence of different auto-ignition modes in a non-uniform reactive field.
Although there are many studies on detonation \cite{Kagan2003,Oran2007,Liberman2010,Kuznethov2010}, supernova \cite{Khokhlov1997,Branch1998,Howell2011,Wang2012,Nonaka2012}, and knocking phenomena \cite{Sankaran2005,Ju2011,Dai2015a,Terashima2015a,Nagano,Morii2021}, most of them mainly addressed the local explosion, i.e., ignition, and very few studies have been conducted in the viewpoint of the interaction between ignition and deflagration.
Recent representative studies that have investigated the relation between "ignition and deflagration" include a theoretical study on the effect of Lewis number and ignition energy on the transition from ignition to flame propagation \cite{Chen2011},
and auto-ignition assisted flames in which burning velocity is increased by low-temperature oxidation in the preheating zone \cite{Ju2019,Zhang2020,Gong2021,Ju2021}. 
These studies have provided in-depth insights into the effect of ignition on the flame propagation.

The objective of this study is to provide general insights into the relation between "ignition and deflagration" in the context of "explosive transition of deflagration." 
At first, the relation between 0D homogeneous ignition and 1D laminar premixed flame with spatial-temporal transformation was theoretically derived.
Next, the effect of Lewis number on the laminar premixed flame structure was addressed in terms of the relation between "ignition and deflagration." 
Finally, to investigate the validity of the present theory to general relation between "ignition and deflagration", computations were performed for hydrogen and $n$-heptane using detailed and reduced kinetics.

\section{Theory}
\label{theory}
\subsection{Governing equations of 0D homogeneous ignition using normalized fuel mass fraction and temperature}
The governing equations for 0D homogeneous ignition are the conservation of energy and chemical species.
For the constant pressure and enthalpy case, the conservation equations are given by
\begin{equation}
  \frac{\mathrm{d}Y_k}{\mathrm{d}t}
  = \frac{\dot{\omega}_k W_k}{\rho} \ \ \ (k = 1, 2, \cdots, K),
  \label{eq:0d-mass-fractions}
\end{equation}
\begin{equation}
  \frac{\mathrm{d}T}{\mathrm{d}t}
  = -\frac{\sum_{k=1}^{K} \dot{\omega}_kh_k W_k}{c_p \rho},
  \label{eq:0d-temperature}
\end{equation}
where $t$ is time, $Y_k$ is the mass fraction of $k^\mathrm{th}$ species,
$\dot{\omega}_k$ is the chemical production rate of $k^\mathrm{th}$ species,
$W_k$ is the molecular weight of $k^\mathrm{th}$ species,
$\rho$ is the mass density,
$T$ is the temperature,
$K$ is the total number of chemical species,
$h_k$ is the enthalpy of $k^\mathrm{th}$ species,
and $c_p$ is the mean specific heat.
The equation of state used for the interconversion of pressure, density, and temperature is a perfect gas given by
$\rho = \frac{p\bar{W}}{RT}$,
where $p$ is the pressure, $\bar{W}$ is the mean molecular weight, and $R$ is the universal gas constant.

\subsection{Legendre transformation}
Consider the Legendre transformation from the multivariate function $f$,
which describe a property of a gas mixture with temperature, density, mass fraction, and time as variables.
The total differential of $f(T, \rho, \bm{Y}, t)$ is given by
\begin{eqnarray*}
  &&\mathrm{d}f(T, \rho, \bm{Y}, t) =\\
  &&\left(\frac{\partial f}{\partial T} \right) \mathrm{d}T
  + \left(\frac{\partial f}{\partial \rho} \right) \mathrm{d}\rho
  + \sum_{k=1}^{K}\left(\frac{\partial f}{\partial Y_k} \right) \mathrm{d} Y_k
  + \left(\frac{\partial f}{\partial t} \right) \mathrm{d}t.
\end{eqnarray*}

Let $a = \left(\partial f / \partial T \right)$ and consider the new multivariate function $g$ given by $g = aT - f$.
The total differential of $g$ is defined by
\begin{equation}
  \mathrm{d}g =
  T\mathrm{d}a
  - \left(\frac{\partial f}{\partial \rho} \right) \mathrm{d}\rho
  - \sum_{k=1}^{K}\left(\frac{\partial f}{\partial Y_k} \right) \mathrm{d}Y_k
  - \left(\frac{\partial f}{\partial t} \right) \mathrm{d}t.
  \label{eq:dg1}
\end{equation}
Thus, $g$ is the function with $a$, $\rho$, $\bm{Y}$, and $t$ as variables.
The total differential of $g(a, \rho, \bm{Y},t)$ is given by
\begin{equation}
  \mathrm{d}g =
  \left(\frac{\partial f}{\partial a} \right) \mathrm{d}a
  + \left(\frac{\partial g}{\partial \rho} \right) \mathrm{d}\rho
  + \sum_{k=1}^{K}\left(\frac{\partial g}{\partial Y_k} \right) \mathrm{d}Y_k
  + \left(\frac{\partial g}{\partial t} \right) \mathrm{d}t.
  \label{eq:dg2}
\end{equation}

Comparing the last terms of Eq. (\ref{eq:dg1}) and Eq. (\ref{eq:dg2}), the relation between $f$ and $g$ is given by
\begin{equation}
  \left(\frac{\partial f}{\partial t}\right) = - \left(\frac{\partial g}{\partial t}\right).
  \label{eq:fandg1}
\end{equation}

Note that $g$ can also be defined by $g = f - aT$.
Then, the relation between $f$ and $g$ is given by
\begin{equation}
  \left(\frac{\partial f}{\partial t}\right) = \left(\frac{\partial g}{\partial t}\right).
  \label{eq:fandg2}
\end{equation}

In Eqs. (\ref{eq:fandg1}) and (\ref{eq:fandg2}), a candidate for $f$ is the normalized temperature $\tilde{T}\in [0,1]$.
A candidate for $g$ could be the normalized fuel mass fraction $\tilde{Y}_\mathrm{f} \in [0,1]$ in the case of single-step chemical reaction models in equation (\ref{eq:fandg1}) or the normalized progress variable $\tilde{C} \in [0, 1]$ in the case of multi-step chemical reaction models in equation (\ref{eq:fandg2}) is considered.
Suppose that once $f$ or $g$ at a certain time is determined, the variables $T$, $\rho$, and $\bm{Y}$ representing the properties of the gas mixture are all determinable.
In other words, assuming that $f$ and $g$ depend only in the time direction, Eqs. (\ref{eq:fandg1}) and (\ref{eq:fandg2}) can transform the partial differential equation into an ordinary differential equation.
As a result, if the normalized temperature $\tilde{T}$ is applied as $f$ and is the normalized fuel mass fraction $\tilde{Y}_\mathrm{f}$ applied as $g$,
the Eq. (\ref{eq:fandg1}) is given by
\begin{equation}
  \frac{\mathrm{d}\tilde{T}}{\mathrm{d}t} = -\frac{\mathrm{d}\tilde{Y}_\mathrm{f}}{\mathrm{d}t}.
  \label{eq:independentTime1}
\end{equation}
Here, the normalized temperature and the normalized fuel mass fraction are defined by
\begin{equation}
  \tilde{Y}_\mathrm{f}
  = \frac{Y_\mathrm{f, 1} - Y_\mathrm{f}}
  {Y_\mathrm{f, 1} - Y_\mathrm{f, 0}}\ \mathrm{and} \ 
  \tilde{T}
  = \frac{T - T_\mathrm{0}}{T_\mathrm{1} - T_\mathrm{0}},
\end{equation}
where subscript $0$ means the initial value and subscript $1$ means the final value.
In addition, if the normalized temperature $\tilde{T}$ is applied as $f$ and the progress variable $\tilde{C}$ for multi-step chemical reaction models is applied as $g$, the Eq. (\ref{eq:fandg2}) is given by
\begin{equation}
  \frac{\mathrm{d}\tilde{T}}{\mathrm{d}t} = \frac{\mathrm{d}\tilde{C}}{\mathrm{d}t}.
  \label{eq:independentTime2}
\end{equation}

Note that, for the Legendre transformation to be possible, $\tilde{T}$, $\tilde{Y}_\mathrm{f}$, and $\tilde{C}$ must be convex functions,
and from Eqs. (\ref{eq:independentTime1}) and (\ref{eq:independentTime2}), $\tilde{T}\in [0,1]$, $\tilde{Y}_\mathrm{f} \in [0,1]$, and $\tilde{C} \in [0, 1]$,
the following conditions are required
$\mathrm{d}\tilde{T}/\mathrm{d}t > 0$, $\mathrm{d}\tilde{Y}_\mathrm{f}/\mathrm{d}t < 0$, and $\mathrm{d}\tilde{C}/\mathrm{d}t > 0$.
This constraint is always true for single-step chemical reaction models and is usually true for multi-step chemical reaction models.
For the sake of simplicity, we will use $\tilde{Y}_\mathrm{f}$ as $g$ in the following discussion, but if you want to use $\tilde{C}$, just reverse the sign.
Note that $\tilde{T}$ and $\tilde{Y}_\mathrm{f}$ are bijective, and once either $\tilde{T}$ or $\tilde{Y}_\mathrm{f}$ is determined, all remaining variables, including T, $\rho$, and $\bm{Y}$, are determined.

%Note that ${Y_\mathrm{f, 0} > Y_\mathrm{f, 1}}$ and ${T_\mathrm{0} < T_\mathrm{1}}$.
The Eqs. (\ref{eq:0d-mass-fractions})(\ref{eq:0d-temperature}) can be transformed using the normalized fuel mass fraction and temperature as follows
\begin{equation}
  \frac{\mathrm{d}\tilde{Y}_\mathrm{f}}{\mathrm{d}t}
  = -\frac{\dot{\omega}_\mathrm{f}W_\mathrm{f}}{\rho(Y_\mathrm{f,1} - Y_\mathrm{f,0})}
  = -\frac{1}{Y_\mathrm{f,1} - Y_\mathrm{f,0}}\frac{\mathrm{d}Y_\mathrm{f}}{\mathrm{d}t},
  \label{eq:0d-mass-fractions-tilde}
\end{equation}
\begin{equation}
  \frac{\mathrm{d}\tilde{T}}{\mathrm{d}t}
  = -\frac{\sum_{k=1}^{K} \dot{\omega}_kh_k W_k}{c_p \rho(T_\mathrm{1} - T_\mathrm{0})}
  = \frac{1}{T_1 - T_0}\frac{\mathrm{d}T}{\mathrm{d}t}.
  \label{eq:0d-temperature-tilde}
\end{equation}

\subsection{Governing equations of 0D homogeneous ignition using residence time}
From Eq. (\ref{eq:independentTime1}), the relation between $\tilde{Y}_\mathrm{f}$ and $\tilde{T}$ is independent of time.
Therefore, Eq. (\ref{eq:independentTime1}) can be transformed using the residence time as follows
\begin{equation}
  \frac{\mathrm{d}\tilde{T}}{\mathrm{d}\tau} = -\frac{\mathrm{d}\tilde{Y}_\mathrm{f}}{\mathrm{d}\tau}.
  \label{eq:independentTau}
\end{equation}
The residence time is the total time that the fluid parcel has spent inside a control volume defined by
\begin{equation}
  \tau = \int_{x_0}^{x} \frac{1}{u} \mathrm{d}x,
  \label{eq:ResidenceTime}
\end{equation}
where $u$ is the velocity of the fluid parcel and $x$ is the position of the fluid parcel.
Then, the total differential of the residence time can be given by
\begin{equation}
  \mathrm{d}\tau = \frac{1}{u} \mathrm{d}x.
  \label{eq:totalDifferential}
\end{equation}

The Eqs. (\ref{eq:0d-mass-fractions-tilde}) and (\ref{eq:0d-temperature-tilde}) can be transformed using the Eq. (\ref{eq:totalDifferential}) as follows
\begin{equation}
  \rho u \frac{\mathrm{d}\tilde{Y}_\mathrm{f}}{\mathrm{d}x}
  = -\frac{\dot{\omega}_\mathrm{f}W_\mathrm{f}}{Y_\mathrm{f,1} - Y_\mathrm{f,0}},
  \label{eq:0d-mass-fractions-tilde-residence}
\end{equation}
\begin{equation}
  \rho u \frac{\mathrm{d}\tilde{T}}{\mathrm{d}x}
  = -\frac{\sum_{k=1}^{K} \dot{\omega}_kh_k W_k}{c_p (T_\mathrm{1} - T_\mathrm{0})}.
  \label{eq:0d-temperature-tilde-residence}
\end{equation}

Therefore, the Eq. (\ref{eq:independentTau}) can be rewritten using the position of fluid parcel as follows
\begin{equation}
  \frac{\mathrm{d}\tilde{T}}{\mathrm{d}x}
  = - \frac{\mathrm{d}\tilde{Y}_\mathrm{f}}{\mathrm{d}x}.
  \label{eq:independentX}
\end{equation}
Note that when Eq. (\ref{eq:independentX}) holds, the relation below also holds
\begin{equation}
  \frac{\sum_{k=1}^{K} \dot{\omega}_kh_k W_k}{c_p (T_\mathrm{1} - T_\mathrm{0})}
  = -\frac{\dot{\omega}_\mathrm{f}W_\mathrm{f}}{Y_\mathrm{f,1} - Y_\mathrm{f,0}}.
  \label{eq:rhs}
\end{equation}

\subsection{Governing equations of 1D laminar premixed flame using normalized fuel mass fraction and temperature}
The conservation equations for 1D laminar premixed flame are given by
\begin{equation}
  \rho u = \mathrm{constant},
\end{equation}
\begin{equation}
  \rho u \frac{\mathrm{d}Y_k}{\mathrm{d}x}
  = \frac{\mathrm{d}}{\mathrm{d}x}\left(\rho D_k\frac{\mathrm{d}Y_k}{\mathrm{d}x}\right)
  + \dot{\omega}_k W_k  \ \ \ (k = 1, 2, \cdots, K),
  \label{eq:1d-mass-fractions}
\end{equation}
\begin{equation}
  \rho u\frac{\mathrm{d}T}{\mathrm{d}x}
  = \frac{1}{c_p}\frac{\mathrm{d}}{\mathrm{d}x}\left(\lambda \frac{\mathrm{d}T}{\mathrm{d}x}\right)
  - \frac{\sum_{k=1}^{K} \dot{\omega}_kh_k W_k}{c_p},
  \label{eq:1d-temperature}
\end{equation}
where $D_k$ is the mixture diffusion coefficient of $k^\mathrm{th}$ species, and $\lambda$ is the thermal conductivity of the mixture.
In this study, we ignore the Soret and Dufour effects, the reasons for this will be justified in section 2.F.

The Eqs. (\ref{eq:1d-mass-fractions}) and (\ref{eq:1d-temperature}) can be rewritten using the normalized fuel mass fraction and temperature as follows
\begin{equation}
  \rho u \frac{\mathrm{d}\tilde{Y}_\mathrm{f}}{\mathrm{d}x}
  = \frac{\mathrm{d}}{\mathrm{d}x}\left(\rho  D_\mathrm{f}\frac{\mathrm{d}\tilde{Y}_\mathrm{f}}{\mathrm{d}x}\right)
  - \frac{\dot{\omega}_\mathrm{f} W_\mathrm{f}}{Y_\mathrm{f,1} - Y_\mathrm{f,0}},
  \label{eq:1d-mass-fractions-tilde}
\end{equation}
\begin{equation}
  \rho u \frac{\mathrm{d}\tilde{T}}{\mathrm{d}x}
  = \frac{1}{c_p}\frac{\mathrm{d}}{\mathrm{d}x}\left(\lambda \frac{\mathrm{d}\tilde{T}}{\mathrm{d}x}\right)
  - \frac{\sum_{k=1}^{K} \dot{\omega}_kh_k W_k}{c_p(T_\mathrm{1} - T_\mathrm{0})}.
  \label{eq:1d-temperature-tilde}
\end{equation}

\subsection{Relation between 0D homogeneous ignition and 1D laminar premixed flame}
Substituting Eqs. (\ref{eq:independentX}) and (\ref{eq:rhs}) into Eq. (\ref{eq:1d-mass-fractions-tilde}) yields the following equation
\begin{equation}
  \rho u \frac{\mathrm{d}\tilde{T}}{\mathrm{d}x}
  = \frac{\mathrm{d}}{\mathrm{d}x}\left(\rho  D_\mathrm{f}\frac{\mathrm{d}\tilde{T}}{\mathrm{d}x}\right)
  - \frac{\sum_{k=1}^{K} \dot{\omega}_kh_k W_k}{c_p(T_\mathrm{1} - T_\mathrm{0})}.
  \label{eq:1d-mass-fractions-final}
\end{equation}
The only difference between Eqs. (\ref{eq:1d-temperature-tilde}) and (\ref{eq:1d-mass-fractions-final}) is the first term of the right-hand side.
Now consider the Lewis number of the fuel defined by $Le_\mathrm{f} = \frac{\lambda}{\rho c_p D_\mathrm{f}}$.
When $Le_\mathrm{f} = 1$, Eqs. (\ref{eq:1d-temperature-tilde}) and (\ref{eq:1d-mass-fractions-final}) are equivalent.
In other words, when $Le_\mathrm{f} = 1$, a 0D homogeneous ignition using normalized fuel mass fraction and temperature is equivalent to a 1D laminar premixed flame using normalized fuel mass fraction and temperature.

\subsection{Lewis number effect on 1D laminar premixed flame using normalized fuel mass fraction and temperature}
Figure 1 shows the relation between 0D homogeneous ignition and 1D laminar premixed flames for hydrogen, methane, propane, and SNIa \cite{Aspden2011}.

\begin{figure}[htbp]
  \begin{center}
    \includegraphics[width=72mm,bb=0 0 567 425]{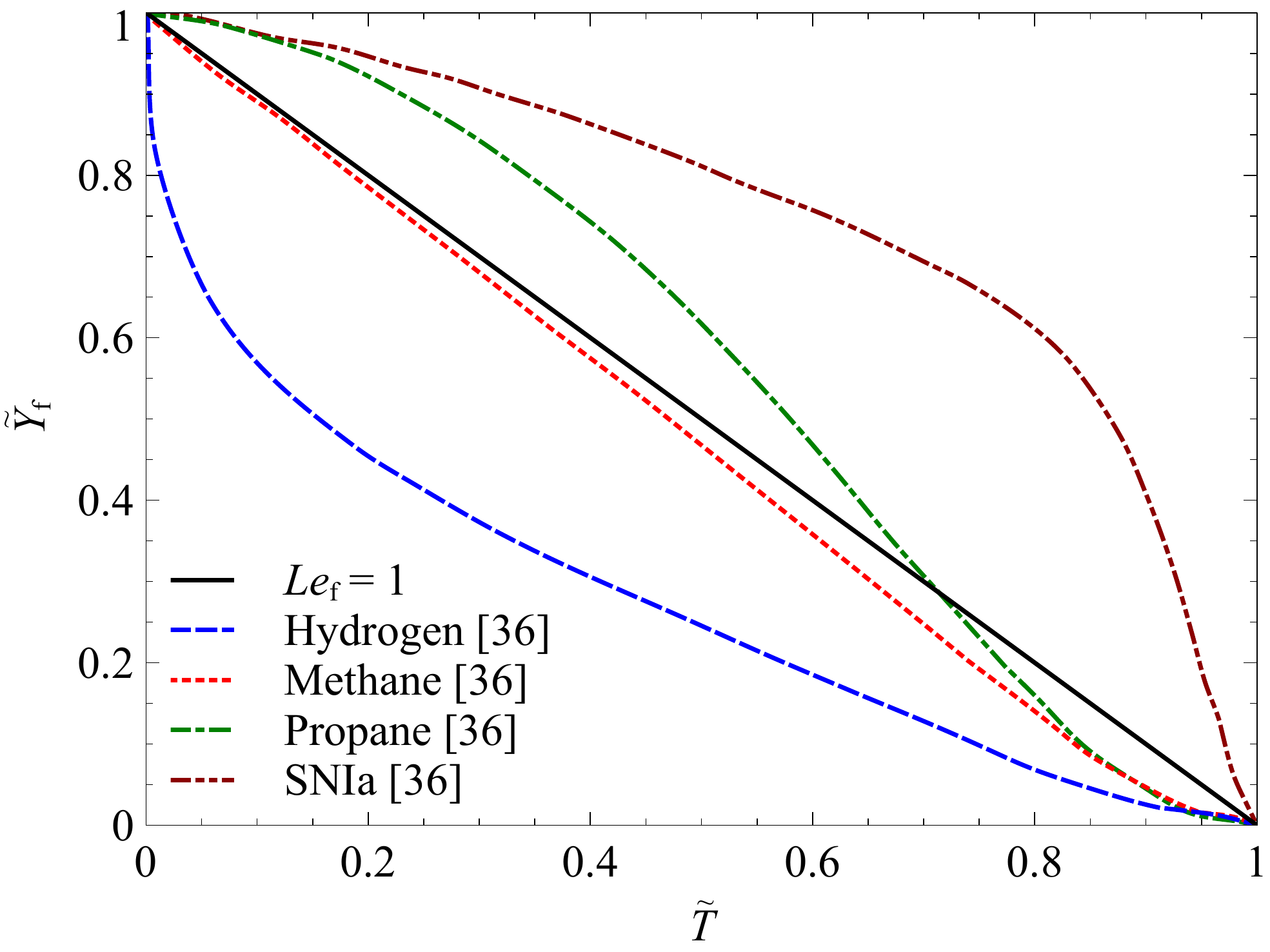}
    \caption{ The relation between 0D homogeneous ignition and 1D laminar premixed flames for hydrogen, methane, propane, and SNIa \cite{Aspden2011} using normalized fuel mass fraction and normalized temperature.
      Note that the 0D homogeneous ignition is equivalent to the 1D laminar premixed flame when $Le_\mathrm{f} = 1$.}
  \end{center}
\end{figure}

In Fig. 1, the SNIa have very high Lewis numbers, and hydrogen, methane, and propane flames have Lewis numbers of 0.36, 0.96 and 1.95.
As discussed in subsection 2.E, for $Le_\mathrm{f} = 1$, 0D homogeneous ignition using normalized fuel mass fraction and temperature is equivalent to 1D laminar premixed flame using normalized fuel mass fraction and temperature.
As described in the paper \cite{Aspden2011},
for $Le_\mathrm{f} < 1$, the profile of a 1D laminar premixed flame is convex below the profile of $Le_\mathrm{f} = 1$ case (i.e., 0D homogeneous ignition),
while for $Le_\mathrm{f} > 1$ the profile is convex above the profile of $Le_\mathrm{f} = 1$ case (i.e., 0D homogeneous ignition) and can intersect the profile of $Le_\mathrm{f} = 1$ case (i.e., 0D homogeneous ignition) in the high-temperature region of the flame (i.e. reaction zone) where the heat release is high.
This is due to the fact that, for $Le_\mathrm{f} < 1$, the diffusion coefficient of the fuel is larger than the thermal conductivity and the fuel is preferentially diffused rather than heated,
while, for $Le_\mathrm{f} > 1$, the thermal conductivity of the fuel is larger than the diffusion coefficient and the fuel being preferentially heated rather than diffused \cite{Aspden2011}.
The influence of the Lewis number on the relation between temperature and fuel mass fraction is discussed theoretically in Buckmater's book \cite{Buckmaster1982}, and the results in Fig. 1 are consistent with that theory.
From now on, the discussion will be conducted in the relatively low temperature region (i.e., preheat zone) where the influence of the Lewis number is clear in Fig. 1.
Thus, in constructing the theory, it can be assumed that the Soret and Dufour effects are negligible.

In Fig. 1, at a given normalized temperature, the consumption of the normalized fuel mass fraction can be considered as the progress of the reaction.
Thus, the reaction progresses are in the order of $Le_\mathrm{f} < 1$, $Le_\mathrm{f} = 1$ (i.e., 0D homogeneous ignition), and $Le_\mathrm{f} > 1$.
Therefore, if $Le_\mathrm{f} < 1$, there is no auto-ignition in the preheat zone.
In other words, 1D laminar premixed flame structure can exist even if the inlet temperature and pressure are very high.
However, if the $Le_\mathrm{f} > 1$, there is a possibility of auto-ignition in the preheat zone.
In other words, depending on the inlet temperature and pressure conditions, there may be conditions where 1D laminar premixed flame structure can not exist.

For example, since the Lewis number of gasoline fuel is greater than unity, knocking can be expected to occur in SI engines whenever the unburned gas region rises to temperature and pressure conditions that make flame propagation impossible.
The Lewis number may also affect the transition from deflagration to detonation, since the characteristic time of the flame behind the shock wave is of the same order of magnitude as that at ignition.
Similarly, since Lewis number is over unity for SNIa, there is likely to be a region behind the shock wave where flame propagation is impossible under some conditions, which may affect the results.

In the next section, simple calculations will be performed to investigate the effect of the Lewis number on the limit of the 1D laminar premixed flame.

\section{Results and discussions}
\label{results}
Computations of 0D homogeneous ignition and 1D laminar premixed flame with multi-step chemical reaction models were performed using Cantera with constant pressure and enthalpy \cite{Cantera}.
The two mixtures, hydrogen/air ($Le<1$) and $n$-heptane/air ($Le>1$), were used.
The multi-step chemical reaction model for hydrogen fuel was UT-JAXA model \cite{Shimizu2011}, and the model for $n$-heptane fuel was reduced SIP model \cite{Sakai2017a, Sakai2018}.
Note that, in order to maintain the Legendre transform even when multi-step chemical reaction models are used, the computational domain should be as short as possible because $\tilde{T}$ and $\tilde{Y}_\mathrm{f}$ must be the convex functions.
Numerical conditions for pressure and equivalent ratio were fixed at 0.1 MPa and 1.0, respectively.
1D laminar premixed flame simulations were performed by varying the inlet temperature from 300 to 3000 K in 100 K increments.
After confirming that the simulation failed no matter how much the computational domain was decreased, the next step was to perform the simulation by varying the inlet temperature in 1K increments to determine the temperature limit at which the calculation could be performed.
After performing 1D laminar premixed flame simulations, 0D homogeneous ignition simulations were performed using the initial temperatures at which the 1D laminar premixed flame could be calculated.

Figure 2 shows the results of burning velocity with varying the inlet temperature.
In the case of hydrogen fuel ($Le _\mathrm{f}<1$), the burning velocity increases with increasing inlet or initial temperature and propagating flame exist even at inlet temperature of 3000 K.
In the case of $n$-heptane fuel ($Le_\mathrm{f}>1$), the burning velocity increases with increasing inlet or initial temperature, but the propagating flame could not exist over 1270 K.
This result is as expected by our theory.

\begin{figure}[htbp]
  \begin{center}
    \includegraphics[width=72mm,bb=0 0 567 425]{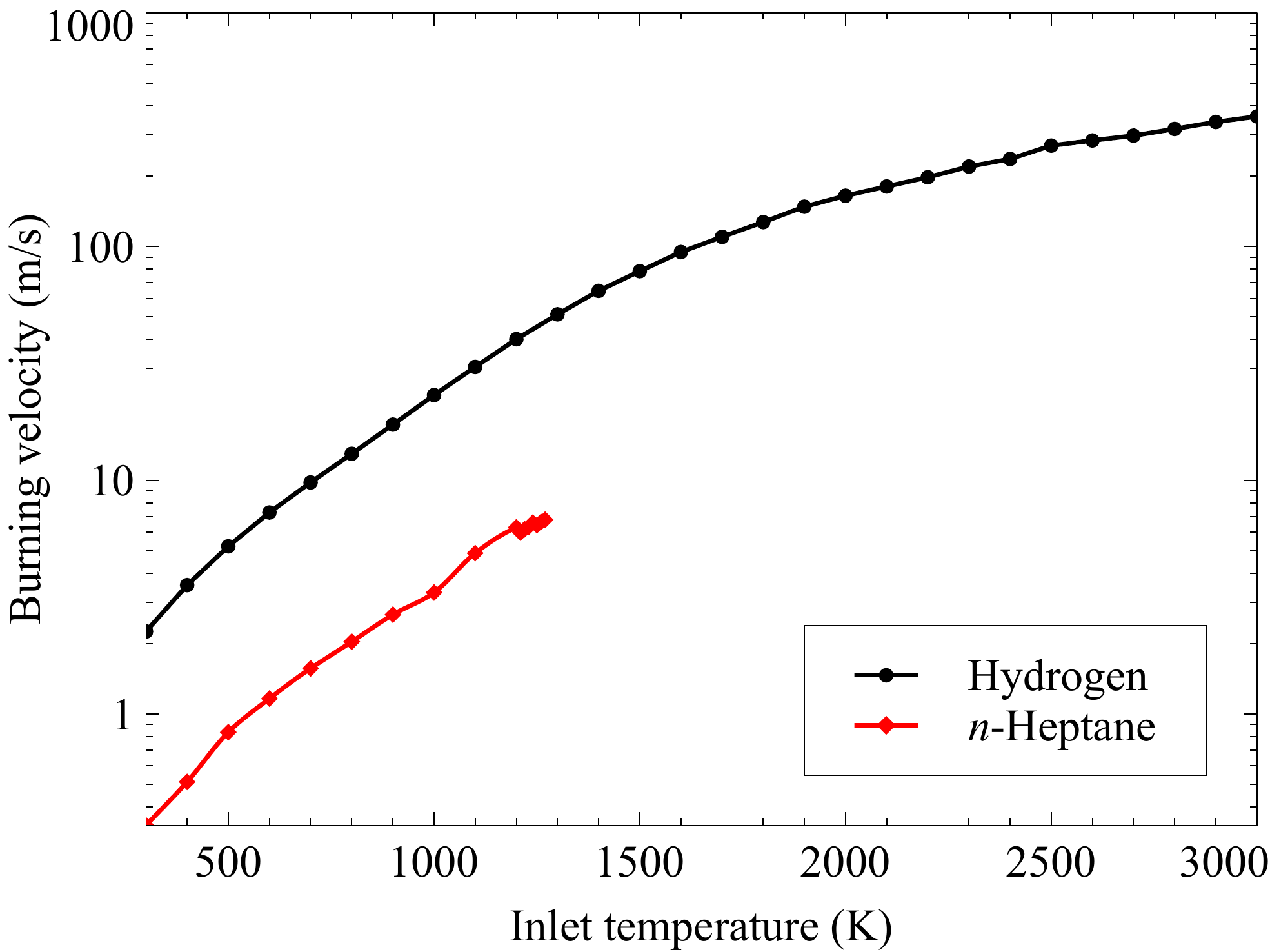}
    \caption{
      Burning velocities with varying inlet temperatures for hydrogen fuel and $n$-heptane fuel with the pressure fixed at 0.1 MPa.
    }
  \end{center}
\end{figure}

Figure 3 shows the temperature time histories of 0D homogeneous ignition and 1D laminar premixed flame for the hydrogen and $n$-heptane fuel cases.
Note that the 1D laminar premixed flame was transformed in the time direction by a spatial-temporal transformation using the residence time in Eq. (\ref{eq:ResidenceTime}).
From Fig. 3(a), it can be seen that 1D laminar premixed flame can be simulated, although the temperature profile of the 0D homogeneous ignition intersects that of the 1D laminar premixed flame under conditions where the initial temperature is above 1500 K.
However, Fig. 3(b) shows that a 1D laminar premixed flame cannot be simulated if the characteristic times of the 0D homogeneous ignition and the 1D laminar premixed flame are of the same order of magnitude.
In other words, in the case of $n$-heptane fuel with $Le_\mathrm{f}>1$, ignition occurs in the preheat zone and a 1D flame structure can not be maintained.

In summary, Lewis number determines the existence of premixed flame structure near the explosive transition of deflagration.

\begin{figure}[htbp]
  \begin{center}
    \includegraphics[width=72mm,bb=0 0 567 425]{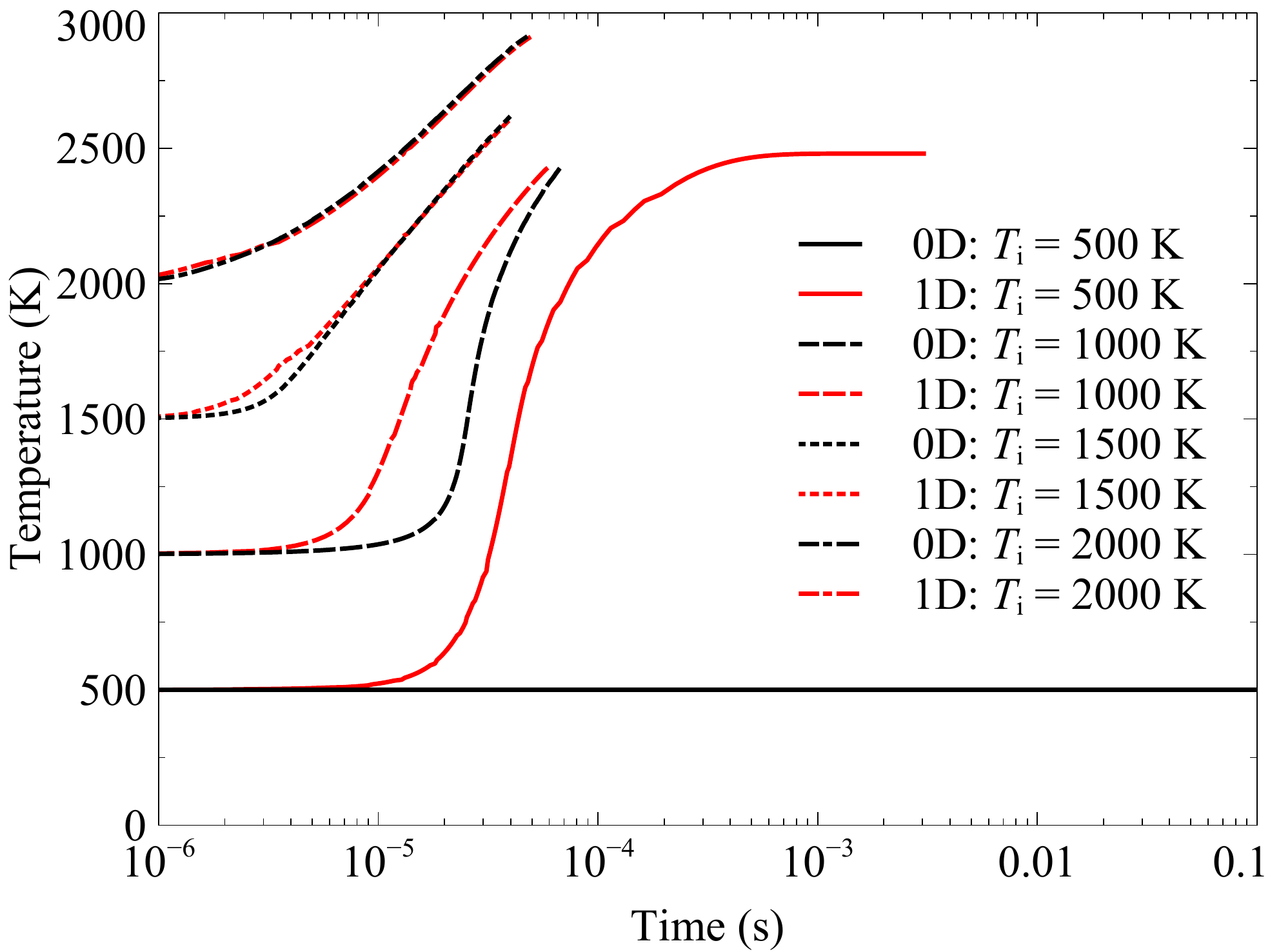} \\
    (a) Hydrogen  \\
    \includegraphics[width=72mm,bb=0 0 567 425]{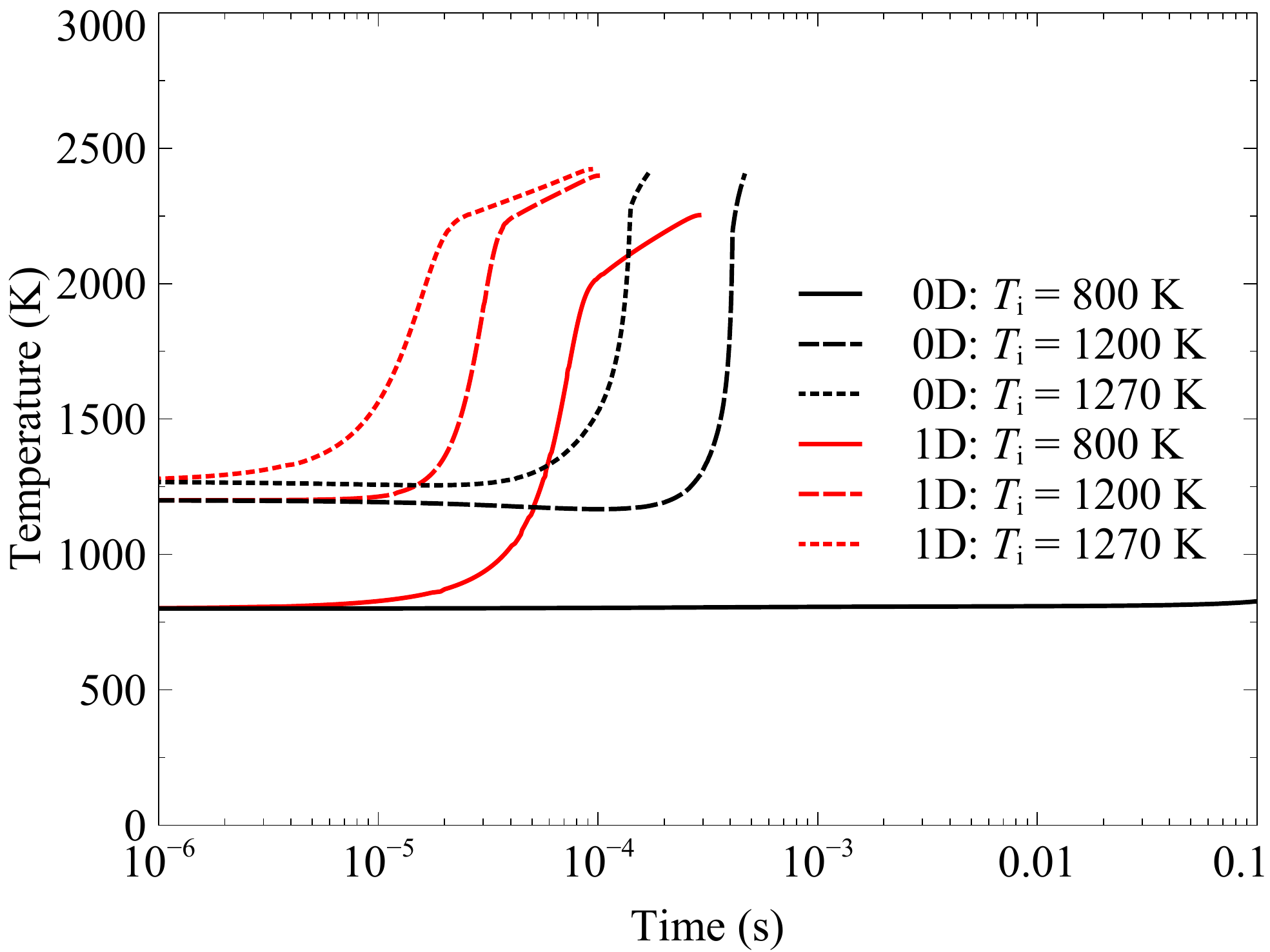} \\
    (b)  $n$-Heptane \\
    \caption{
      Time history of temperature with varying $T_i$ for (a) hydrogen fuel and (b) $n$-heptane fuel, with the pressure fixed at 0.1 MPa.
      $T_i$ is initial temperature in 0D simulations and is the inlet temperature in 1D simulations.
      The simulations could be performed even at $T_i = 3000$ K for hydrogen fuel, but not for $n$-heptane fuel above $T_i = 1270$ K.
    }
  \end{center}
\end{figure}

\section{Conclusions}
\label{conclusions}
It is theoretically shown that the relation between the normalized fuel mass fraction and the normalized temperature is the same for 0D homogeneous ignition and 1D laminar premixed flame with $Le_\mathrm{f}=1$ after a spatial-temporal transformation.
Also, for 1D laminar premixed flames with $Le_\mathrm{f}<1$, no ignition occurs in the preheat zone, so the flame structure is always present.
In the case of 1D laminar premixed flames with $Le_\mathrm{f} > 1$, there is a possibility of ignition in the preheat zone, and no flame structure exists when the temperature is higher than a certain threshold.
In other words, if the characteristic times of ignition and deflagration are about the same order of magnitude, such as behind a shock wave or in a near explosive transition of SI engine, ignition and deflagration are connected and discussed via the Lewis number.

\section*{Acknowledgement}
This work was partially supported by JSPS KAKENHI Grant Number 19KK0097.

\bibliography{main}% Produces the bibliography via BibTeX.

\end{document}